\documentclass[aps,pra,twocolumn,a4paper,amssymb,amsmath,footinbib,showpacs]{revtex4}
\usepackage{revsymb}
\usepackage{multirow}
\usepackage{graphicx}
\usepackage{subfigure}
\usepackage{graphics}
\usepackage{psfrag}

\newcommand{\ket}[1]{| #1 \rangle}

\begin{document}

\title{Polynomial invariants and Bell inequalities  \\
as entanglement measure of 4-qubit states}
\author{Jochen \surname{Endrejat}}
\email[Electronic address:]{ jochen.endrejat@uni-bayreuth.de}
\author{Helmut \surname{B\"uttner}}
\affiliation{Theoretische Physik I, Universit{\"a}t Bayreuth, 
  D-95440 Bayreuth, Germany}
\date{\today}
\begin{abstract}
We compare the polynomial invariants for four qubits introduced by
Luque and Thibon, PRA {\bf 67}, 042303 (2003), with optimized Bell inequalities
and a combination of two qubit concurrences. It is shown for various parameter dependent
states from different SLOCC classes that it is possible to measure 
a genuine 4-qubit entanglement with these polynomials.
\end{abstract}
\maketitle

The classification of multiqubit entanglement is 
actually a wideley discussed field. Until now no measure
for genuine $n$-qubit entanglement ($n \ge 4$) is known,
in contrast to the 2-qubit concurrence or the 3-qubit tangle.
In this paper we will show by example that the combination of 
optimized Bell inequalities and polynomial invariants, introduced
by Luque and Thibon \cite{Luque:03}, yields an entanglement measure
for four qubits.\\
In \cite{Endrejat:04} we argued that the comparison between 
an optimization of Bell-type inequalities \cite{Yu:03} and a combination
of the global entanglement measure $Q$ \cite{MeyerWallach:02} with the
sum of the squared 2-qubit concurrences $C_{ij}^2$ \cite{Hill:97,Wootters:98} 
yields the same parameter dependence of the entanglement measures.
We will show that one of the invariants shows the same behavior.\\ 
The states which are in the focus of our research belong to different
entanglement classes. Miyake \cite{Miyake:03} and Verstraete et al. \cite{Verstraete:02}
discussed the SLOCC classification of multiqubit states. Miyake 
connected the classification to hyperdeterminants and showed that the 
representative of the outermost 4-qubit entanglement class is the state 
\begin{multline}
\ket{G_{\alpha \beta \gamma \delta}} =
\alpha \bigl(\ket{0000}+\ket{1111}\bigr) + \beta \bigl(\ket{0011} + \ket{1100}\bigr)\\
+\gamma \bigl(\ket{0101} + \ket{1010}\bigr) + \delta \bigl(\ket{0110} + \ket{1001}\bigr)
\end{multline}
This state is equivalent to the state $G_{abcd}$ of
Vertraetes nine SLOCC families for four qubits. A criterion for this class is a
nonzero hyperdeterminant $Det A_4$, called $\Delta$ in the paper by Luque and Thibon.\\ 
For the calculation of the SLOCC class affiliation, we take
the easy-to-use criteria recently introduced by Li et al. \cite{Li:06}.
There starting points were representative states like the GHZ- or the W-state and an
existence criterion for the SLOCC transformation.
Two states $\ket{\phi}$, $\ket{\psi}$ are invariant under SLOCC  if
local invertible $2 \times 2$ matrices $A, B, C, D$ exist with
$\ket{\phi} = A \otimes B \otimes C \otimes D \ket{\psi}$.
Taking the state under consideration as $\ket{\phi}$ and the final state
as $\ket{\psi}$, one can easily calculate
these criteria, that means equation under the condition that the 
matrices exist.\\
In the following we discuss the results for three different classes of states.
\\
{\it 1)} In our first example we will consider the 4-qubit parameter dependent
GHZ state, 
\begin{equation}
\gamma \ket{0000}+\sqrt{1-\gamma^2}\ket{1111}
\end{equation}
with $\gamma \in ]0,1[$. 
{
\psfrag{g}[][][0.8]{$\gamma^2$}
\psfrag{yeins}[][][0.8]{Concurrences / Invariants}
\psfrag{yzwei}[][][0.8]{Optimized Inequalities}
\begin{figure}[b]
\begin{center}
\includegraphics[width=7cm]{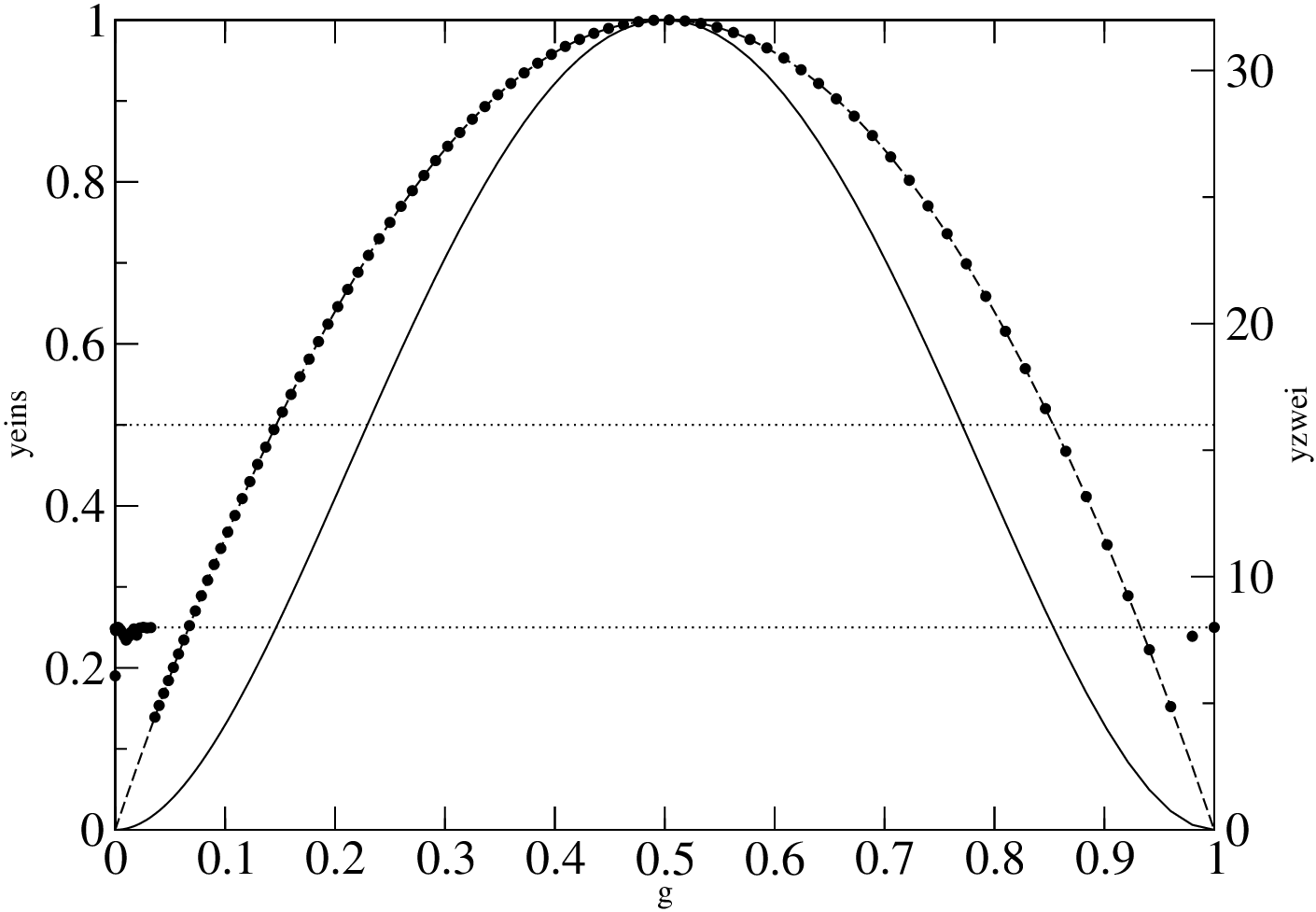}
\caption{\label{plot-ghz}Optimized Bell inequality (black dots) on the right
y-axis, Global Entanglement $Q$ (dashed line) 
and $S$ invariant (black line) on the left y-axis, as function of $\gamma^2$ for
the 4-qubit parameter dependent GHZ state. The dotted lines show the Bell inequality 
condition for 3-qubit entanglement ($\ge 8$) resp. 4-qubit entanglement ($\ge 16$).}
\end{center}
\end{figure}
}
It is assumed that this state shows genuine 4-qubit entanglement. The 
concurrences between any pair of qubits, $C_{ij}$, are 0.
They are calculated the usual way \cite{Hill:97,Wootters:98}.
The concurrence is defined as the maximum 
$C_{ij}=\text{max}\{\sqrt{\lambda_1}-\sqrt{\lambda_2}-\sqrt{\lambda_3}-\sqrt{\lambda_4},0\}$,
and the $\lambda_i$ are the eigenvalues of the matrix
$\rho_{ij}=\rho_{ij}(\sigma^y\otimes\sigma^y)\rho_{ij}^*(\sigma^y\otimes\sigma^y)$,
with $\rho_{ij}$, the reduced density matrix to the qubits $i$ and $j$.\\
The global entanglement is $Q=4 \gamma^2(1-\gamma^2)$.
The calculation of the Luque/Thibon invariants yields $H=\gamma \sqrt{1-\gamma^2}$,
$L=M=N=0$, $D_{xt}=0$, $S= \gamma^4(-1+\gamma^2)^2 /12$, 
$T=- \gamma^6(-1+\gamma^2)^3 /216$ and $\Delta=0$ (see appendix). It is nicely seen
that the $S$ resp. $T$ invariant and the global entanglement measure are the same,
up to a square root and a constant factor. In Fig.\ref{plot-ghz} we show the Global Entanglement $Q$,
the invariant $S$ and the Bell optimization as function of $\gamma^2$. \\
We use the Mermin-Klyshko-type Bell inequalities. You get the corresponding operators out
of the following recursion relations:
\begin{align}
F_N    &= \frac{1}{2}(D_N + D_N')F_{N-1} + \frac{1}{2}(D_N - D_N')F'_{N-1} \label{fn}\\
F_{N}' &= \frac{1}{2}(D_N + D_N')F_{N-1}' + \frac{1}{2}(D_N' - D_N)F_{N-1} \label{fns}
\end{align}
with $F_2 = (A'B + AB')+( AB -A'B') $ and $F'_2 =(A'B + AB') -( AB -A'B')$.
The $A^{(\prime)}, B^{(\prime)}$ resp. $D^{(\prime)}$  can be written as sums of Pauli matrices, e.g.
$A^{(\prime)} = \vec{a}^{(\prime)} \cdot \vec{\sigma_A}$ for qubit $A$, with normalized
vectors $\vec{a}^{(\prime)}$. These $F_N^{(\prime)}$ operators are optimized and yield
a criterion for 4-qubit entanglement, if 
$\langle F_4 \rangle ^2 + \langle F'_4 \rangle ^2 > 16$ \cite{Yu:03,Endrejat:04}.\\
Though the Bell optimization for the parameter dependent GHZ state
has some difficulties near $\gamma \sim 0$ and $\gamma \sim 1$ as described in \cite{Endrejat:04},
the three quantities match nicely, especially quantified at the maximum $\gamma^2 = 1/2$.\\
The state does not belong to the outermost 4-qubit SLOCC class as described by Miyake,
because the hyperdeterminant criterion is not fulfilled, $\Delta=0$. The simple criteria for 
the GHZ class after Li et al.is $-\gamma \sqrt{1-\gamma^2} \neq 0$ and is fullfilled 
for $\gamma \in ]0,1[$.\\ 
\\
{\it 2)} The classification of the next state is more complex. Parts of it were 
already done in \cite{Endrejat:04}. The state is one of the eigenstates of a special
4-qubit Heisenberg model, with antiferromagnetic coupling constants $J$ and $J_s$ bot $\ge 0$,
and is written as:
\begin{multline}
\ket{\phi_2} = \beta_1 \bigl(-\ket{0011} + \ket{0110} - \ket{1001} + \ket{1100}\bigr)\\
 +\beta_2 \bigl( -\ket{0101} + \ket{1010} \bigr)  
\end{multline}
with the two amplitudes $\beta_1$ and $\beta_2$ given by
$\beta_1 = \bigl(4 + (-J + 2 J_s + \delta)^2/(2 J^2)\bigr)^{-1/2}$, 
$\beta_2 = \bigl(4+(-J + 2 J_s + \delta)^2/(2 J^2)\bigr)^{1/2} J/(2 \delta) $
and the abbreviation $\delta = (9 J^2 -4 J J_s + 4 J_s^2)^{1/2}$.\\
{
\psfrag{J}[][][0.8]{$J, J_s =2$}
\psfrag{Js}[][][0.8]{$J_s, J=2$}
\psfrag{yeins}[][][0.8]{Concurrences / Invariants}
\psfrag{yzwei}[][][0.8]{Optimized Inequalities}
\begin{figure}
\begin{center}
\subfigure{\includegraphics[width=7cm]{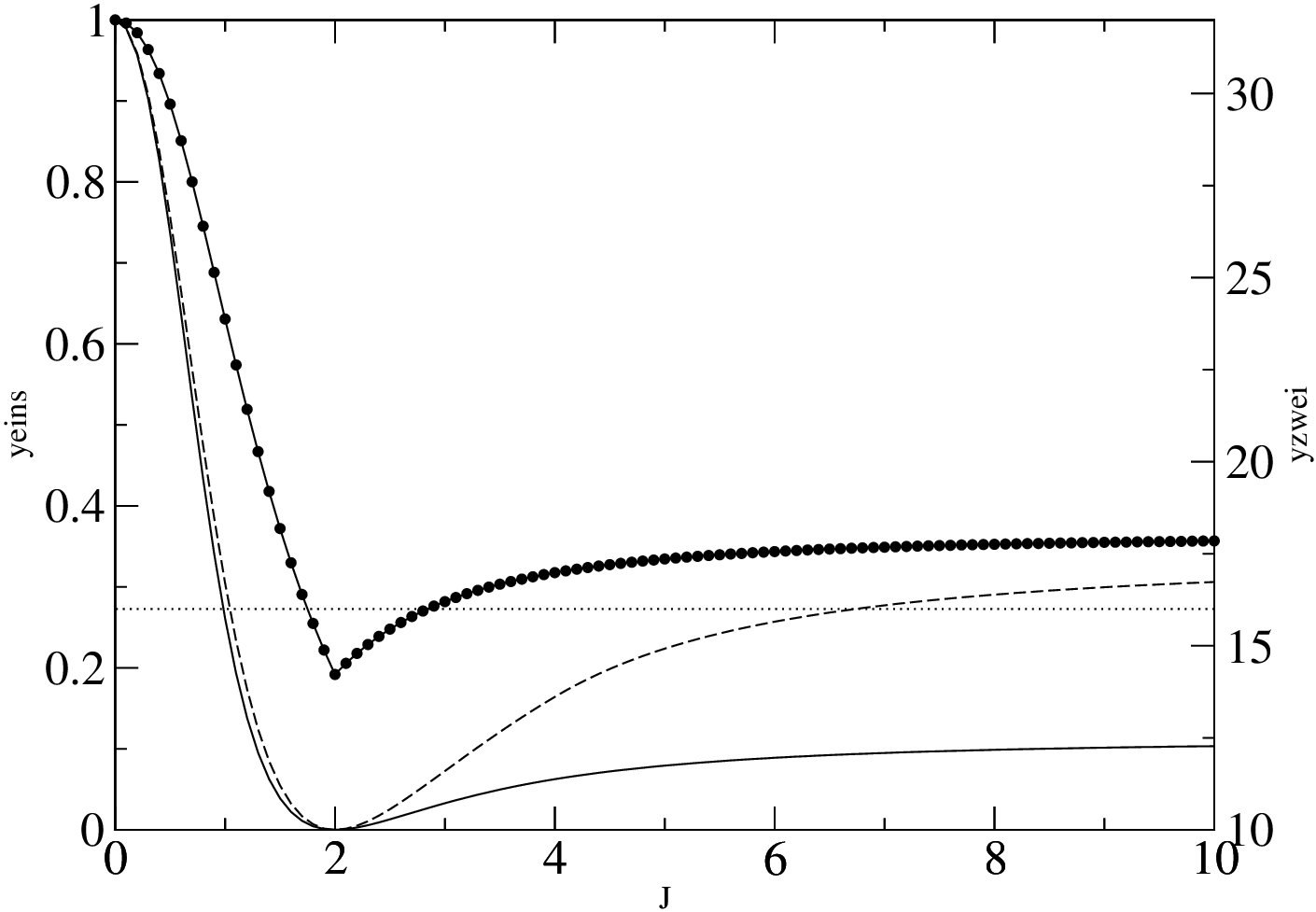}}
\hfill
\subfigure{\includegraphics[width=7cm]{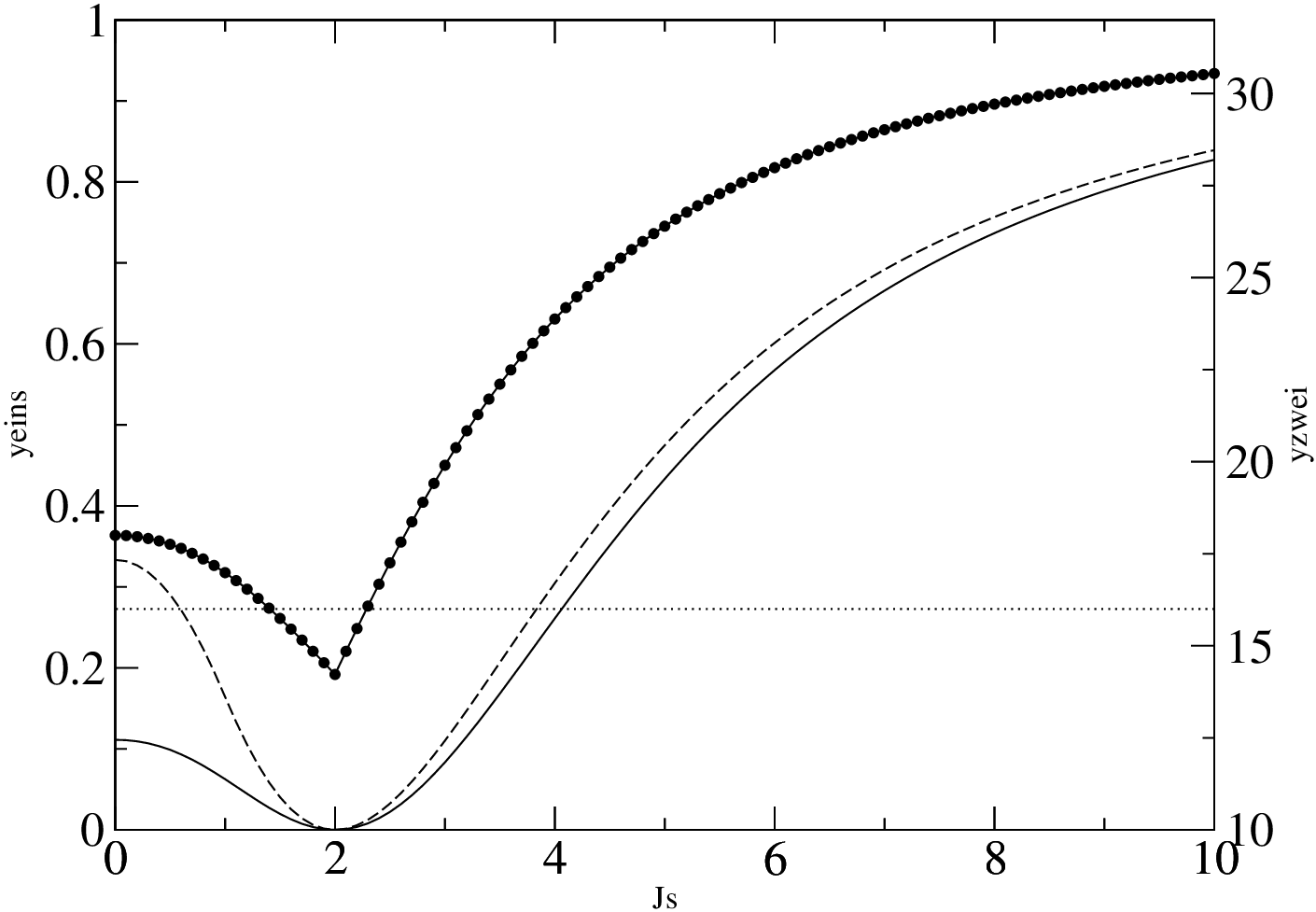}}
\caption{\label{plot-phi2} Optimized Bell inequality (line with dots) on the right y-axis,
$1-\sum C_{ij}^2$ (dashed line) and $S$ invariant (black line) as function 
of $J$ resp. $J_s$ for the state $\ket{\phi_2}$ on the left y-axis.
The dotted line shows the Bell inequality 
condition for 4-qubit entanglement. $J$ and $J_s$ are complicated functions of 
$\beta_1$ and $\beta_2$ (see text).}
\end{center}
\end{figure}
}
We know from the optimization of Bell-type inequalities that this state
fullfills the criteria for 3- resp. 4-qubit entanglement. The global
entanglement measure is constant, $Q=1$. Additionally the
parameter dependence of the optimized inequality is similar to the course of the combination 
$Q-\sum C_{ij}^2 \equiv 1-\sum C_{ij}^2 $. The parameter dependence of both
is shown in Fig. \ref{plot-phi2}, as a function of $J$ with constant $J_s=2$ resp.
of $J_s$ with constant $J=2$.\\
The calculation of the Luque/Thibon invariants yields the following.
$H = -2 \beta_1^2 - \beta_2^2$, $L = -N = \beta_1^2 (\beta_1^2 -\beta_2^2)$,
$M = 0$ and $D_{xt} = -\beta_1^2 \beta_2^4$ (see appendix). The more interesting ones are 
\begin{align}
S &= \frac{1}{12} \beta_2^4 (-4 \beta_1^2 + \beta_2^2)^2\\
T &= \frac{1}{216} \beta_2^6 (-4 \beta_1^2 + \beta_2^2)^3
\end{align}
and $\Delta = 0$, with the general relation $\Delta=S^3-27T^2$. In
Fig. \ref{plot-phi2} we also show the paramater dependence of the invariant $S$,
which is normalized to 1. 
All three quantities plotted show the same behavior. The coincidence is very nice seen
at the maxima resp. minima. For $J=0, J_s=2$ resp. $J=2, J_s \to \infty$ 
all three quantities show a maximal 4-qubit entanglement. 
Also the minimum at $J=2, J_s=2$ matches very nicely.\\
Since the invariant $\Delta$ is equal to 0, the state does not belong to the 
$G_{\alpha \beta \gamma \delta}$ class. But the state fullfills  for almost all 
paramters, except a small range around $J=J_s=2$,the Bell condition for  4-qubit entanglement.
Because of this fact we will test the classification criteria for the GHZ SLOCC class as introduced 
by Li et al. \cite{Li:06}. We get the following equations:
\begin{equation}
2\beta_1^2 + \beta_2^2  \neq  0, \quad
-\beta_1^4 = 0 \quad \text{and} \quad \beta_1^2 \beta_2^2 = 0.
\end{equation} 
These are solved for $\beta_1 =0$ and $\beta_2 \neq 0$.
This is the case for the limits $J_s=2,J\to 0$ and $J=2, J_s \to \infty$.
The state is reduced to a GHZ type state $\ket{\phi_2}=(-\ket{0101}+\ket{1010})/\sqrt{2}$,
and belongs to the GHZ SLOCC class. In all other cases, that means $\beta_1 \neq 0$,
the state belongs to another not yet quantified SLOCC class.\\
\\
{
\psfrag{gamma}[][][0.8]{$\gamma$}
\psfrag{yeins}[][][0.8]{Concurrences / Invariants}
\psfrag{yzwei}[][][0.8]{Optimized Inequalities}
\begin{figure}[t]
\begin{center}
\includegraphics[width=7cm]{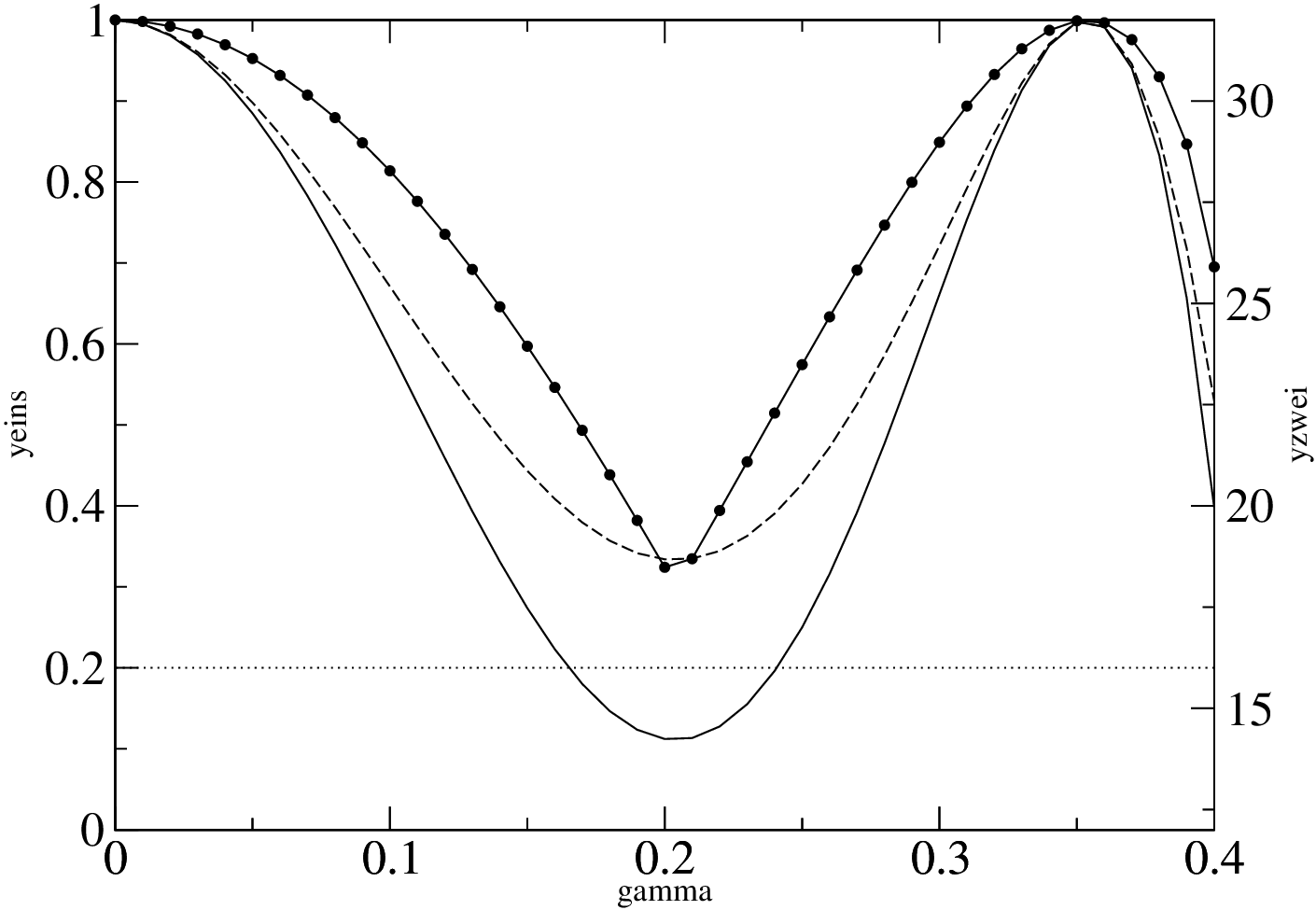}
\caption{\label{plot-miyake}Optimized Bell inequality (line with dots) on the right
y-axis, $1-\sum C_{ij}^2$ (dashed line) 
and $S$ invariant (black line) as function of $\gamma$ for the state 
$\ket{G_{\alpha \gamma}}$, on the left y-axis. The dotted line shows the Bell inequality 
condition for 4-qubit entanglement.}
\end{center}
\end{figure}
}
{\it 3)} In the next part we study a state with $\Delta \neq 0$. Therefore we
take the Miyake state $\ket{G_{\alpha \beta \gamma \delta}}$ and reduce the number of paramters.
We set $\beta=\delta=\gamma$ and $2\alpha^2 + 6\gamma^2=1$:
\begin{multline}
\ket{G_{\alpha \gamma}} =
\alpha \bigl(\ket{0000}+\ket{1111}\bigr) + \gamma \bigl(\ket{0011} + \ket{1100}+\\
\ket{0101} + \ket{1010} + \ket{0110} + \ket{1001}\bigr),
\end{multline}
and we choose $\alpha$ resp. $\gamma$ real, $\gamma \in [0,1/\sqrt{6}]$.
Again we calculate the Luque/Thibon invariants and compare them with the
Bell optimization and the combination of global entanglement 
and squared concurrences. The global entanglement is constant $Q=1$ and 
the concurrences are all equal,
$C_{12}=C_{13}=C_{14}=C_{23}=C_{24}=C_{34}$, with
\begin{equation}
C_{12}=
\begin{cases}
4\gamma(\alpha-\gamma)  &\text{if}\quad 0 \le \gamma \le \frac{1}{2\sqrt{2}}\\
2 (\gamma^2 -\alpha^2) &\text{if}\quad \frac{1}{2\sqrt{2}} < \gamma \le \frac{1}{\sqrt{6}}      
\end{cases}
\end{equation}
with $\alpha$ defined above.\\ 
For the Luque/Thibon invariants we get the following:
$H=1/2$, $L=M=N=0$, $D_{xt}= 1/4\gamma^2 (1-8 \gamma^2)^2$ and
\begin{align}
S &= \frac{1}{192} -\frac{1}{4}\gamma^2(1-8\gamma^2)^2\\
T&=\frac{1}{13824}-\frac{\gamma^2}{192}+\frac{7\gamma^4}{48}-\frac{7\gamma^6}{3}\\
&\qquad+24\gamma^8-128\gamma^{10}+256\gamma^{12}\\
\Delta &= -\frac{1}{512} (6\gamma^2-1)(24\gamma^2-1)^2(8\gamma^3 -\gamma)^6
\end{align}
In Fig. \ref{plot-miyake} we show the parameter dependence of our measures.
It is again clearly seen that the course of the invariant $S$, which is again normalized
to 1, the optimized inequality and the function of the squared concurrences yield the 
same result in the parameter dependence. Especially for $\gamma=0$, the state reduces to a GHZ
state, and all three quantities have a maximum. Also the minimum at $\gamma=1/2\sqrt{2}$ and the
maximum at $\gamma = 1/2\sqrt{6}$ match.\\
The state $\ket{G_{\alpha \beta \gamma \delta}}$ is the representative of the outermost
SLOCC entanglement class with the criteria $\Delta \neq 0$. Also for our special choosen 
state $\ket{G_{\alpha \gamma}}$ the hyperdeterminant $\Delta$ is different from 0, except for three values,
$\gamma = \pm 1/\sqrt{6},\pm 1/\sqrt{8},\pm 1/\sqrt{24}$. We now take the criteria for the 
GHZ SLOCC class by Li et al. 
\begin{equation}
-\alpha^2 - 3\gamma^2 \neq  0 \quad \wedge \quad
\alpha^2 \gamma^2 - \gamma^4 =0
\end{equation}
and test them with the calculated roots for $\Delta$. For $\gamma=1/\sqrt{8}$ these equations
are solved and the state belongs to the GHZ class.\\
\\
Another interesting feature 
of the invariants is found from the comparison
of the invariant $H$ and the inequality which comes out of the criteria for the 
GHZ SLOCC class. Up to a minus sign, they are equal in all our examples. That means,
a nonzero invariant $H$ is a criterion for the affiliation of a state to the GHZ SLOCC
class.\\
\\
{\it Conclusions and discussions -}
It is shown in this paper that the genuine 4-qubit entanglement could be
measured with a polynomial invariant. This invariant $S$ yields the same
parameter dependence as optimized Bell-type inequalities and a combination
of global entanglement and 2-qubit concurrences for the states we have choosen
from different SLOCC classes.
\appendix*
\section{Luque/Thibon invariants}
Here we give the general equation for the 
Luque/Thibon invariant $D_{xt}$, beacause it is not explicitly calculated in 
\cite{Luque:03}. A pure 4-qubit state can be written in the computational basis
as $\ket{\psi}=\sum_{i=0}^{15} a_i \ket{i}$.
The invariants can then be expressed in terms of the $a_i$. Here especially the
invariant $D_{xt}$:
\begin{multline}
D_{xt}=(-a_{11} a_{13} + a_{15} a_{9}) 
(-(a_{3} a_{4} + a_{2} a_{5} - a_{1} a_{6} - a_{0} a_{7})\\ 
(-a_{0} a_{14} + a_{12} a_{2} + a_{10} a_{4} - a_{6} a_{8}) + 
(a_{2} a_{4} - a_{0} a_{6})\\
(-a_{1} a_{14} - a_{0} a_{15} + a_{13} a_{2} + a_{12} a_{3} +\\ 
a_{11} a_{4} + a_{10} a_{5} - a_{7} a_{8} - a_{6} a_{9})) +\\
(-a_{10} a_{12} + a_{14} a_{8})(-(a_{3} a_{5} - a_{1} a_{7}) \\
(-a_{1} a_{14} - a_{0} a_{15} + a_{13} a_{2} + a_{12} a_{3} +\\
 a_{11} a_{4} + a_{10} a_{5} - a_{7} a_{8} - a_{6} a_{9}) +\\
 (a_{3} a_{4} + a_{2} a_{5} - a_{1} a_{6} - a_{0} a_{7})\\
 (-a_{1} a_{15} + a_{13} a_{3} + a_{11} a_{5} - a_{7} a_{9})) -\\ 
(-a_{11} a_{12} - a_{10} a_{13} + a_{15} a_{8} + a_{14} a_{9})
((a_{3} a_{5} - a_{1} a_{7}) \\
(a_{0} a_{14} - a_{12} a_{2} - a_{10} a_{4} + a_{6} a_{8}) +
(-a_{2} a_{4} + a_{0} a_{6}) \\
(a_{1} a_{15} - a_{13} a_{3} - a_{11} a_{5} + a_{7} a_{9}))
\end{multline}


\end{document}